\documentclass[prl,twocolumn,nofootinbib,floatfix,superscriptaddress]{revtex4-1}

\usepackage[utf8]{inputenc}
\usepackage[T1]{fontenc}
\usepackage[pdftex]{graphicx}
\usepackage{amsmath, amstext, amssymb, amsfonts, amsxtra}
\usepackage[tbtags]{mathtools}
\usepackage{textcomp}
\usepackage{xspace}
\usepackage{icomma}
\usepackage{dcolumn}
\usepackage{hyperref}

\newcommand{\ket}[1]{\ensuremath{\vert #1 \rangle}\xspace}%
\newcommand{\avg}[1]{\ensuremath{\langle #1 \rangle}\xspace}%
\newcommand{\kb}{\ensuremath{k_\textsc{b}}\xspace}%
\newcommand{\Er}{\ensuremath{E_\text{r}}\xspace}%
\newcommand{\oned}{1\textsc{d}\xspace}%
\newcommand{\twod}{2\textsc{d}\xspace}%
\newcommand{\lsup}[1]{\ensuremath{{\vphantom{x}}^{#1}}}%
\newcommand{\Rb}{\ensuremath{\lsup{87}\text{Rb}}\xspace}
\DeclareMathOperator*{\Ai}{Ai}%

\begin{document}

\title{Light-cone-like spreading of correlations in a quantum many-body system}

\author{Marc~Cheneau}%
\email[Electronic address: ]{marc.cheneau@mpq.mpg.de}%
\affiliation{Max-Planck-Institut f\"{u}r Quantenoptik$,$ 85748 Garching$,$ Germany}%
\author{Peter~Barmettler}%
\author{Dario~Poletti}%
\affiliation{D\'{e}partement de physique th\'{e}orique$,$ Universit\'{e} de Gen\`{e}ve$,$ 1211
  Gen\`{e}ve$,$ Switzerland}%
\author{Manuel~Endres}%
\author{Peter~Schau\ss}%
\author{Takeshi~Fukuhara}%
\author{Christian~Gross}%
\affiliation{Max-Planck-Institut f\"{u}r Quantenoptik$,$ 85748 Garching$,$ Germany}%
\author{Immanuel~Bloch}%
\affiliation{Max-Planck-Institut f\"{u}r Quantenoptik$,$ 85748 Garching$,$ Germany}%
\affiliation{Ludwig-Maximilians-Universit\"{a}t$,$ 80799 M\"{u}nchen$,$ Germany}%
\author{Corinna~Kollath}%
\affiliation{D\'{e}partement de physique th\'{e}orique$,$ Universit\'{e} de Gen\`{e}ve$,$ 1211
  Gen\`{e}ve$,$ Switzerland}%
\affiliation{Centre de physique th\'{e}orique$,$ \'{E}cole Polytechnique$,$ CNRS$,$ 91128
  Palaiseau$,$ France}%
\author{Stefan~Kuhr}%
\affiliation{Max-Planck-Institut f\"{u}r Quantenoptik$,$ 85748 Garching$,$ Germany}%
\affiliation{University of Strathclyde$,$ SUPA$,$ Glasgow G4 0NG$,$ United Kingdom}%


\begin{abstract}
  How fast can correlations spread in a quantum many-body system?  Based on the seminal work by Lieb
  and Robinson \cite{Lieb:1972}, it has recently been shown that several interacting many-body
  systems exhibit an effective light cone that bounds the propagation speed of correlations
  \cite{Bravyi:2006, Calabrese:2006, Eisert:2006, Nachtergaele:2006a}. The existence of such a
  ``speed of light'' has profound implications for condensed matter physics and quantum information,
  but has never been observed experimentally. Here we report on the time-resolved detection of
  propagating correlations in an interacting quantum many-body system. By quenching a
  one-dimensional quantum gas in an optical lattice, we reveal how quasiparticle pairs transport
  correlations with a finite velocity across the system, resulting in an effective light cone for
  the quantum dynamics. Our results open important perspectives for understanding relaxation of
  closed quantum systems far from equilibrium \cite{Polkovnikov:2011b} as well as for engineering
  efficient quantum channels necessary for fast quantum computations \cite{Bose:2007}.
\end{abstract}

\maketitle

In contrast to relativistic quantum field theory, no ``speed limit'' exists in non-relativistic
quantum mechanics, allowing in principle for the propagation of information over arbitrary distances
in arbitrary short times \cite{Bravyi:2006}. However, one could naively expect that in real physical
systems short-range interactions allow information to propagate only with a finite velocity. The
existence of a maximal velocity, also called Lieb--Robinson bound, has indeed been shown
theoretically in some systems, e.g. interacting spins on a lattice \cite{Bravyi:2006,
  Calabrese:2006, Eisert:2006, Nachtergaele:2006a}, but to which extent this result can be
generalised remains an open question \cite{Lauchli:2008, Nachtergaele:2009a, Cramer:2008c,
  Eisert:2009}. Lieb--Robinson bounds have already found a number of fundamental applications
\cite{Hastings:2004a, Nachtergaele:2011}. For example, they allow for a rigorous proof of a
long-standing conjecture that linked the presence of a spectral gap in a lattice system to the
exponential decay of correlations in the ground state \cite{Nachtergaele:2006b, Hastings:2006}. They
also provide fundamental scaling laws for the entanglement entropy, which is an indicator of the
computational cost for simulating strongly interacting systems \cite{Eisert:2010}.

In the context of quantum many-body systems, the existence of a Lieb--Robinson bound can be probed
by recording the dynamics following a sudden parameter change (quench) in the Hamiltonian. In that
case, a simple picture has been suggested: quantum-entangled quasiparticles emerge from the
initially highly excited state and propagate ballistically \cite{Calabrese:2006}, carrying
correlations across the system. Ultracold atomic gases offer an ideal testbed to explore such
quantum dynamics due to their almost perfect decoupling from the environment and their fast
tunability \cite{Bloch:2008}. In addition, the recently demonstrated technique of single-site
imaging in an optical lattice \cite{Bakr:2009, Sherson:2010} offers the resolution and sensitivity
necessary to reveal the dynamical evolution of a many-body system at the single-particle level.

\begin{figure}
  \includegraphics[width=1.0\columnwidth]{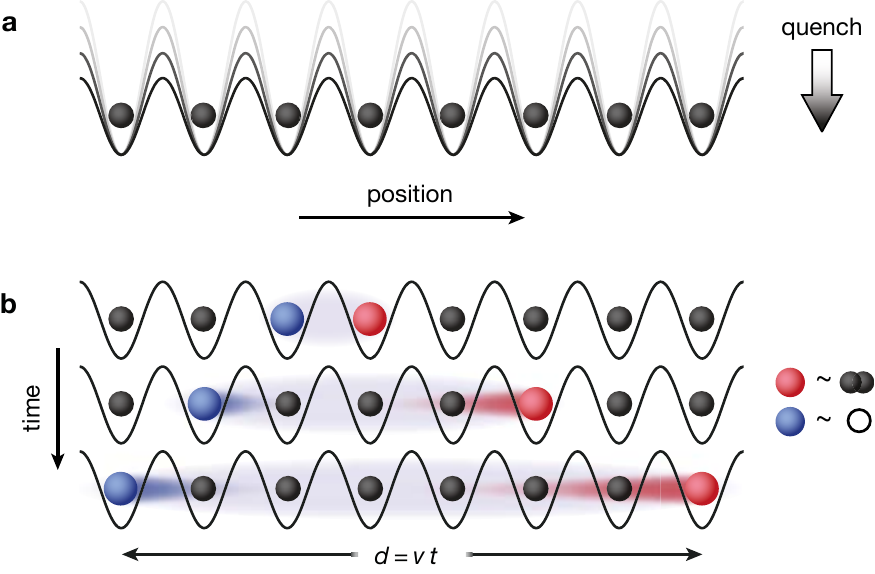}
  \caption{{\bf Spreading of correlations in a quenched atomic Mott insulator.} \textbf{a}, A \oned
    ultracold gas of bosonic atoms (black balls) in an optical lattice is initially prepared deep in
    the Mott-insulating phase with unity filling. The lattice depth is then abruptly lowered,
    bringing the system out of equilibrium. \textbf{b}, Following the quench, entangled
    quasiparticle pairs emerge at all sites. Each of these pairs consists of a doublon (red ball)
    and a holon (blue ball) on top of the unity-filling background, which propagate ballistically in
    opposite directions. It follows that a correlation in the parity of the site occupancy builds up
    at time $t$ between any pair of sites separated by a distance $d = vt$, where $v$ is the
    relative velocity of the doublons and holons.}
  \label{fig:1}
\end{figure}

Our system consists of ultracold bosonic atoms in an optical lattice and is well described by the
Bose--Hubbard model \cite{Fisher:1989, Jaksch:1998a}. This model is parameterised by two energy
scales: the on-site interaction, $U$, and the tunnel coupling between adjacent sites, $J$. Driven by
the competition of these two parameters, a quantum phase transition between a superfluid and a
Mott-insulating phase occurs in homogeneous systems with integer filling $\bar n$. In the
one-dimensional (\oned) geometry considered here, the critical point of this transition is located
at $(U/J)_{\text c} \simeq 3.4$ \cite{Kuhner:2000}. We observed the time evolution of spatial
correlations after a fast decrease of the effective interaction strength $U/J$ from an initial value
deep in the Mott-insulating regime, with filling $\bar n =1$, to a final value closer to the
critical point (Fig.~\ref{fig:1}a). After such a quench, the initial many-body state
$\ket{\Psi_{0}}$ is highly excited and acts as a source of quasiparticles. In order to elucidate the
nature and the dynamics of these quasiparticles, we have developed an analytical model in which the
occupancy of each lattice site is restricted to $n = 0, 1$ or 2 (see Appendix). For large
interaction strengths, the quasiparticles consist of either an excess particle (doublon) or a hole
(holon) on top of the unity-filling background. Fermionizing these quasiparticles with a
Jordan–Wigner transformation allows us to partially eliminate the non-physical states in which a
lattice site would be occupied by two quasiparticles. To first order in $J/U$, we then find that the
many-body state at time $t$ after the quench reads:
\begin{multline}
  \label{eq:1}
  \ket{\Psi(t)} \simeq \ket{\Psi_{0}} + \, i \sqrt{8} \, \frac{J}{U} \sum_{k} \sin(k a_{\text{lat}})
  \\ \, \Big[ 1- e^{-i [\epsilon_{\text{d}}(k) + \epsilon_{\text{h}}(-k) ]t/\hbar} \Big] \, \hat
  d^\dagger_{k} \, \hat h_{-k}^{\dagger} \ket{\Psi_{0}} \; ,
\end{multline}
with $a_{\text{lat}}$ the lattice period. Here $\hat d^{\dagger}_{k}$ and $\hat h_{k}^{\dagger}$ are
the creation operators for a doublon and a holon with momentum $k$, respectively, and $k$ belongs to
the first Brillouin zone. Quasiparticles thus emerge at any site in the form of entangled pairs,
consisting of a doublon and a holon with opposite momenta. Some of these pairs are bound on
nearest-neighbour sites while the others form wave packets, due to their peaked momentum
distribution. The wave packets propagate in opposite directions with a relative group velocity $v$
determined by the dispersion relation $\epsilon_{\text{d}}(k)+\epsilon_{\text{h}}(-k)$ of doublons
and holons (Fig.~\ref{fig:1}b). The propagation of quasiparticle pairs is reflected in the two-point
parity correlation functions \cite{Endres:2011}:
\begin{equation}
  \label{eq:2}
  C_{d}(t) = \avg{\hat s_{j}(t) \hat s_{j+d}(t)} - \avg{\hat s_{j}(t)} \avg{\hat s_{j+d}(t)}\; ,
\end{equation}
where $j$ labels the lattice sites. The operator $\hat s_{j}(t) = e^{i\pi [\hat n_{j}(t) - \bar n
  ]}$ measures the parity of the occupation number $\hat n_{j}(t)$. It yields +1 in the absence of
quasiparticles (odd occupancy) and -1 if a quasiparticle is present (even occupancy). Because the
initial state is close to a Fock state with one atom per lattice site, we expect $C_{d}(t=0) \simeq
0$. After the quench, the propagation of quasiparticle pairs with the relative velocity $v$ results
in a positive correlation between any pair of sites separated by a distance $d = vt$.

The experimental sequence started with the preparation of a two-dimensional (\twod) degenerate gas
of \Rb confined in a single antinode of a vertical optical lattice \cite{Sherson:2010, Endres:2011}
($z$-axis, \mbox{$a_{\text{lat}}=532$\,nm}). The system was then divided into about 10 decoupled
\oned chains by adding a second optical lattice along the $y$-axis and by setting both lattice
depths to $20.0(5)$\,\Er, where $\Er = (2\pi\hbar)^{2}/(8 m a_{\text{lat}}^{2})$ is the recoil
energy of the lattice and $m$ the atomic mass of \Rb. The effective interaction strength along the
chains was tuned via a third optical lattice along the $x$-axis. The number of atoms per chain
ranged between 10 and 18, resulting in a lattice filling $\bar n = 1$ in the Mott-insulating
domain. The inital state was prepared by adiabatically increasing the $x$-lattice depth until the
interaction strength reached a value of $(U/J)_{0} = 40(2)$. At this point, we measured the
temperature to be $T \simeq 0.1\,U/\kb$ (\kb is the Boltzmann constant) following the method
described in Ref. \cite{Sherson:2010}. We then brought the system out of equilibrium by lowering the
lattice depth typically within $100$\,\textmu s, which is fast compared to the inverse tunnel
coupling $\hbar/J$, but still adiabatic with respect to transitions to higher Bloch bands. The final
lattice depths were in the Mott-insulating regime, close to the critical point. After a variable
evolution time, we ``froze'' the density distribution of the many-body state by rapidly raising the
lattice depth in all directions to $\sim 80$\,\Er. Finally, the atoms were detected by fluorescence
imaging using a microscope objective with a resolution on the order of the lattice spacing and a
reconstruction algorithm extracted the occupation number at each lattice site
\cite{Sherson:2010}. Because inelastic light-assisted collisions during the imaging lead to a rapid
loss of atom pairs, we directly detected the parity of the occupation number.

\begin{figure}
  \includegraphics[width=0.8\columnwidth]{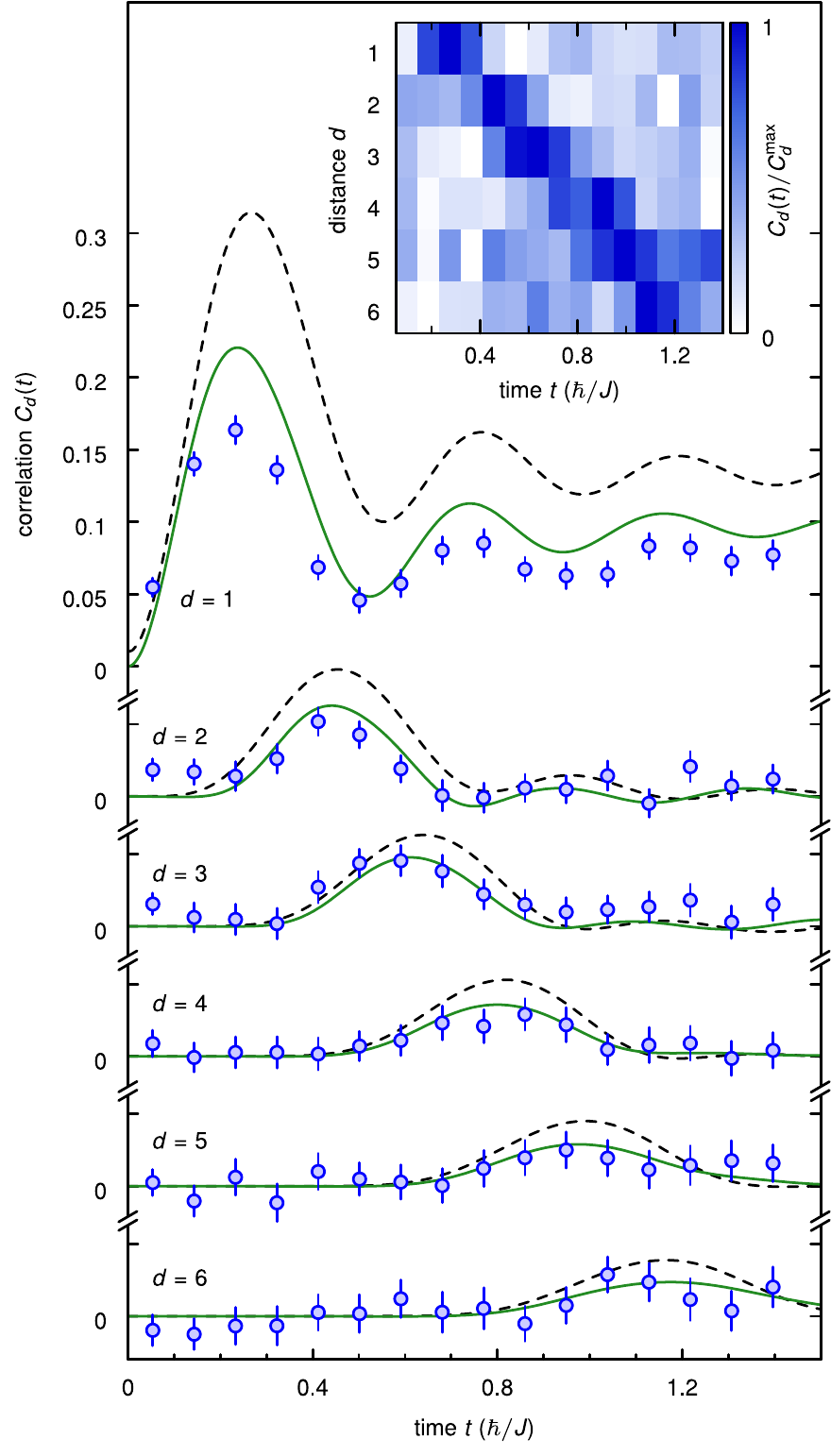}
  \caption{{\bf Time evolution of the two-point parity correlations.} After the quench, a positive
    correlation signal propagates with increasing time to larger distances. The experimental values
    for a quench from $U/J=40$ to $U/J=9.0$ (circles) are in good agreement with the corresponding
    numerical simulation for an infinite, homogeneous system at zero temperature (continuous
    line). Our analytical model (dashed line) also qualitatively reproduces the observed
    dynamics. Inset: Experimental data displayed as a colormap, revealing the propagation of the
    correlation signal with a well defined velocity. The experimental values result from the average
    over the central $N$ sites of more than 1000 chains, where $N$ equals $80\,\%$ of the length of
    each chain. Error bars represent the standard deviation.}
  \label{fig:2}
\end{figure}

Our experimental results for the time evolution of the two-point parity correlations after a quench
to $U/J=9.0(3)$ show a clear positive signal propagating with increasing time to larger distances
$d$ (Fig.~\ref{fig:2}). In addition, the propagation velocity of the correlation signal is constant
over the range $2 \leq d \leq 6$ (inset of Fig.~\ref{fig:2}). We found similar dynamics also for
quenches to $U/J=5.0(2)$ and 7.0(3) (Fig.~\ref{fig:S1}). We note that the observed signal cannot be
attributed to a simple density wave because such an excitation would result in $\avg{\hat s_{j} \hat
  s_{j+d}} = \avg{\hat s_{j}} \avg{\hat s_{j+d}}$. We compared the experimental results to numerical
simulations of an infinite, homogeneous system at $T=0$ using the adaptive time-dependent density
matrix renormalization group \cite{Daley:2004, White:2004} (\emph{t}-DMRG). In the simulation, the
initial and final interaction strengths were fixed at the experimentally determined values and the
quench was considered instantaneous, at $t=0$. We found remarkable agreement between the experiment
and theory over all explored distances and times, despite the finite temperature and the harmonic
confinement with frequency $\nu = 68(1)\,$Hz that characterise the experimental system. The observed
dynamics is also qualitatively reproduced by our analytical model for $U/J=9.0$. For lower values of
$U/J$, however, the model breaks down due to the increasing number of quasiparticles.

\begin{figure}
  \includegraphics[width=0.9\columnwidth]{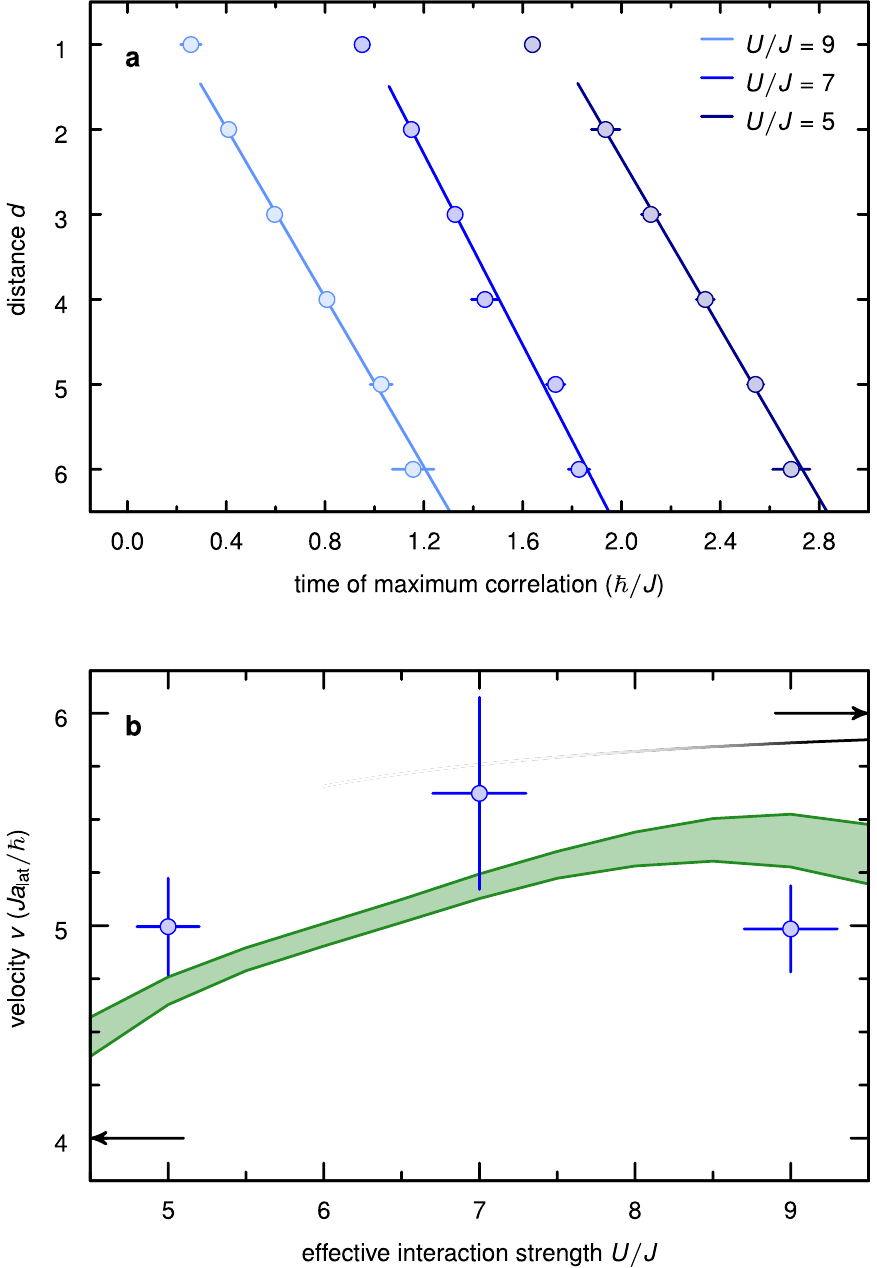}
  \caption{{\bf Propagation velocity.} \textbf{a}, Determination of the propagation velocity for the
    quenches to $U/J=5.0$ (triangles), 7.0 (squares) and 9.0 (circles). The time of the maximum of
    the correlation signal is obtained from fits to the traces $C_{d}(t)$. Error bars represent the
    68\,\% confidence interval of these fits. We then extract the propagation velocities from
    weigthed linear fits restricted to $2 \leq d \leq 6$ (lines). The data for $U/J = 5.0$ and 7.0
    have been offset horizontally for clarity. \textbf{b}, Comparison of the experimental velocities
    (circles) to the ones obtained from numerical simulations for an infinite, homogeneous system at
    zero temperature (shaded area). The shaded area and the vertical error bars denote the 68\,\%
    confidence interval of the fit. The horizontal error bars represent the uncertainty due to the
    calibration of the lattice depth. The black line corresponds to the bound $v_{\text{max}}$
    predicted by our effective model (the fading indicates the break down of this model). The arrows
    mark the maximum velocity expected in the non-interacting case (left) and the asymptotic value
    derived from our model when \mbox{$U/J\to\infty$} (right).}
  \label{fig:3}
\end{figure}

We extracted the propagation velocity $v$ from the time of the correlation peak as a function of the
distance $d$ (Fig.~\ref{fig:3}a). A linear fit restricted to $2 \leq d \leq 6$ yields \mbox{$v
  \times \hbar/(J a_{\text{lat}}) = 5.0(2)$}, $5.6(5)$ and $5.0(2)$ for $U/J = 5.0(2)$, 7.0(3) and
9.0(3), respectively. The points for $d=1$ were excluded from the fit, as they result from the
interference between propagating and bound quasiparticle pairs (see Eq. 1). A comparison of the
experimental velocities with the ones obtained from numerical simulations (Fig.~\ref{fig:3}b) shows
agreement within the error bars. The measured velocities can also be compared with two limiting
cases: On the one hand, they are significantly larger than the spreading velocity of non-interacting
particles, $v=4\,Ja_{\text{lat}}/\hbar$, and twice the velocity of sound in the superfluid phase
\cite{Kollath:2005b}; on the other hand, they remain below the maximum velocity $v_{\text{max}}
\approx (6Ja_{\text{lat}}/\hbar) \left[1 - 16J^2/(9U^2) \right]$ predicted by our analytical model,
that can be interpreted as a Lieb--Robinson bound (Fig.~\ref{fig:3}b). In the limit \mbox{$U/J
  \to\infty$}, this bound corresponds to doublons and holons propagating with the respective group
velocities $4\,Ja_{\text{lat}}/\hbar$ and $2\,Ja_{\text{lat}}/\hbar$. The higher velocity of
doublons simply reflects their Bose-enhanced tunnel coupling.

In conclusion, we have presented the first experimental observation of an effective light cone for
the spreading of correlations in an interacting quantum many-body system. Although the observed
dynamics can be understood within a fermionic quasiparticle picture valid deep in the Mott
insulating regime, we note that our experimental data cover a region close to the critical point,
for which only ab-initio numerical simulations are available so far \cite{Lauchli:2008}. Our work
opens interesting perspectives, such as revealing the entanglement carried by the quasiparticle
pairs or investigating the quantum dynamics in higher dimensions, where little is known about
Lieb--Robinson bounds and the scaling of entanglement. For example, the experiment can be extended
to study correlation propagation in two dimensions, where existing numerical and analytical
approaches suffer from severe limitations. Furthermore, the production rate of excitations and the
domain formation when tuning the effective interaction strength slowly across the critical point can
be investigated, thereby exploring a quantum analog to the Kibble--Zurek mechanism
\cite{Kibble:1976, Zurek:1985, Polkovnikov:2011b}.

\section*{Acknowledgements}

We thank C. Weitenberg and J. F. Sherson for their contribution to the design and construction of
the apparatus. We also thank D. Baeriswyl, T. Giamarchi, V. Gritsev and S. Huber for
discussions. C.K. acknowledges previous collaboration on a related subject with A. Läuchli. We
acknowledge funding by MPG, DFG, EU (NAMEQUAM, AQUTE, Marie Curie Fellowship to M.C.), JSPS
(Postdoctoral Fellowship for Research Abroad to T.F.), ``Triangle de la physique'', ANR (FAMOUS) and
SNSF (under division II). Financial support for the computer cluster on which the calculations were
performed has been provided by the ``Fondation Ernst et Lucie Schmidheiny''.

\bibliography{LightCone.bib}

\begin{thebibliography}{35}%
\makeatletter
\providecommand \@ifxundefined [1]{%
 \@ifx{#1\undefined}
}%
\providecommand \@ifnum [1]{%
 \ifnum #1\expandafter \@firstoftwo
 \else \expandafter \@secondoftwo
 \fi
}%
\providecommand \@ifx [1]{%
 \ifx #1\expandafter \@firstoftwo
 \else \expandafter \@secondoftwo
 \fi
}%
\providecommand \natexlab [1]{#1}%
\providecommand \enquote  [1]{``#1''}%
\providecommand \bibnamefont  [1]{#1}%
\providecommand \bibfnamefont [1]{#1}%
\providecommand \citenamefont [1]{#1}%
\providecommand \href@noop [0]{\@secondoftwo}%
\providecommand \href [0]{\begingroup \@sanitize@url \@href}%
\providecommand \@href[1]{\@@startlink{#1}\@@href}%
\providecommand \@@href[1]{\endgroup#1\@@endlink}%
\providecommand \@sanitize@url [0]{\catcode `\\12\catcode `\$12\catcode
  `\&12\catcode `\#12\catcode `\^12\catcode `\_12\catcode `\%12\relax}%
\providecommand \@@startlink[1]{}%
\providecommand \@@endlink[0]{}%
\providecommand \url  [0]{\begingroup\@sanitize@url \@url }%
\providecommand \@url [1]{\endgroup\@href {#1}{\urlprefix }}%
\providecommand \urlprefix  [0]{URL }%
\providecommand \Eprint [0]{\href }%
\@ifxundefined \urlstyle {%
  \providecommand \doi  [0]{\begingroup \@sanitize@url \@doi}%
  \providecommand \@doi [1]{\endgroup \@@startlink {\doibase
  #1}doi:\discretionary {}{}{}#1\@@endlink }%
}{%
  \providecommand \doi  [0]{doi:\discretionary{}{}{}\begingroup
  \urlstyle{rm}\Url }%
}%
\providecommand \doibase [0]{http://dx.doi.org/}%
\providecommand \Doi [0]{\begingroup \@sanitize@url \@Doi }%
\providecommand \@Doi  [1]{\endgroup\@@startlink{\doibase#1}\@@Doi}%
\providecommand \@@Doi [1]{#1\@@endlink}%
\providecommand \selectlanguage [0]{\@gobble}%
\providecommand \bibinfo  [0]{\@secondoftwo}%
\providecommand \bibfield  [0]{\@secondoftwo}%
\providecommand \translation [1]{[#1]}%
\providecommand \BibitemOpen [0]{}%
\providecommand \bibitemStop [0]{}%
\providecommand \bibitemNoStop [0]{.\EOS\space}%
\providecommand \EOS [0]{\spacefactor3000\relax}%
\providecommand \BibitemShut  [1]{\csname bibitem#1\endcsname}%
\bibitem [{\citenamefont {Lieb}\ and\ \citenamefont
  {Robinson}(1972)}]{Lieb:1972}%
  \BibitemOpen
  \bibfield  {author} {\bibinfo {author} {\bibfnamefont {E.~H.}\ \bibnamefont
  {Lieb}}\ and\ \bibinfo {author} {\bibfnamefont {D.~W.}\ \bibnamefont
  {Robinson}},\ }\Doi {10.1007/BF01645779} {\bibfield  {journal} {\bibinfo
  {journal} {Commun. Math. Phys.},\ }\textbf {\bibinfo {volume} {28}},\
  \bibinfo {pages} {251} (\bibinfo {year} {1972})}\BibitemShut {NoStop}%
\bibitem [{\citenamefont {Bravyi}\ \emph {et~al.}(2006)\citenamefont {Bravyi},
  \citenamefont {Hastings},\ and\ \citenamefont {Verstraete}}]{Bravyi:2006}%
  \BibitemOpen
  \bibfield  {author} {\bibinfo {author} {\bibfnamefont {S.}~\bibnamefont
  {Bravyi}}, \bibinfo {author} {\bibfnamefont {M.~B.}\ \bibnamefont
  {Hastings}}, \ and\ \bibinfo {author} {\bibfnamefont {F.}~\bibnamefont
  {Verstraete}},\ }\Doi {10.1103/PhysRevLett.97.050401} {\bibfield  {journal}
  {\bibinfo  {journal} {Phys. Rev. Lett.},\ }\textbf {\bibinfo {volume} {97}},\
  \bibinfo {pages} {050401} (\bibinfo {year} {2006})}\BibitemShut {NoStop}%
\bibitem [{\citenamefont {Calabrese}\ and\ \citenamefont
  {Cardy}(2006)}]{Calabrese:2006}%
  \BibitemOpen
  \bibfield  {author} {\bibinfo {author} {\bibfnamefont {P.}~\bibnamefont
  {Calabrese}}\ and\ \bibinfo {author} {\bibfnamefont {J.}~\bibnamefont
  {Cardy}},\ }\Doi {10.1103/PhysRevLett.96.136801} {\bibfield  {journal}
  {\bibinfo  {journal} {Phys. Rev. Lett.},\ }\textbf {\bibinfo {volume} {96}},\
  \bibinfo {pages} {136801} (\bibinfo {year} {2006})}\BibitemShut {NoStop}%
\bibitem [{\citenamefont {Eisert}\ and\ \citenamefont
  {Osborne}(2006)}]{Eisert:2006}%
  \BibitemOpen
  \bibfield  {author} {\bibinfo {author} {\bibfnamefont {J.}~\bibnamefont
  {Eisert}}\ and\ \bibinfo {author} {\bibfnamefont {T.~J.}\ \bibnamefont
  {Osborne}},\ }\Doi {10.1103/PhysRevLett.97.150404} {\bibfield  {journal}
  {\bibinfo  {journal} {Phys. Rev. Lett.},\ }\textbf {\bibinfo {volume} {97}},\
  \bibinfo {pages} {150404} (\bibinfo {year} {2006})}\BibitemShut {NoStop}%
\bibitem [{\citenamefont {Nachtergaele}\ \emph {et~al.}(2006)\citenamefont
  {Nachtergaele}, \citenamefont {Ogata},\ and\ \citenamefont
  {Sims}}]{Nachtergaele:2006a}%
  \BibitemOpen
  \bibfield  {author} {\bibinfo {author} {\bibfnamefont {B.}~\bibnamefont
  {Nachtergaele}}, \bibinfo {author} {\bibfnamefont {Y.}~\bibnamefont {Ogata}},
  \ and\ \bibinfo {author} {\bibfnamefont {R.}~\bibnamefont {Sims}},\ }\Doi
  {10.1007/s10955-006-9143-6} {\bibfield  {journal} {\bibinfo  {journal} {J.
  Stat. Phys.},\ }\textbf {\bibinfo {volume} {124}},\ \bibinfo {pages} {1}
  (\bibinfo {year} {2006})}\BibitemShut {NoStop}%
\bibitem [{\citenamefont {Polkovnikov}\ \emph {et~al.}(2011)\citenamefont
  {Polkovnikov}, \citenamefont {Sengupta}, \citenamefont {Silva},\ and\
  \citenamefont {Vengalattore}}]{Polkovnikov:2011b}%
  \BibitemOpen
  \bibfield  {author} {\bibinfo {author} {\bibfnamefont {A.}~\bibnamefont
  {Polkovnikov}}, \bibinfo {author} {\bibfnamefont {K.}~\bibnamefont
  {Sengupta}}, \bibinfo {author} {\bibfnamefont {A.}~\bibnamefont {Silva}}, \
  and\ \bibinfo {author} {\bibfnamefont {M.}~\bibnamefont {Vengalattore}},\
  }\Doi {10.1103/RevModPhys.83.863} {\bibfield  {journal} {\bibinfo  {journal}
  {Rev. Mod. Phys.},\ }\textbf {\bibinfo {volume} {83}},\ \bibinfo {pages}
  {863} (\bibinfo {year} {2011})}\BibitemShut {NoStop}%
\bibitem [{\citenamefont {Bose}(2007)}]{Bose:2007}%
  \BibitemOpen
  \bibfield  {author} {\bibinfo {author} {\bibfnamefont {S.}~\bibnamefont
  {Bose}},\ }\Doi {10.1080/00107510701342313} {\bibfield  {journal} {\bibinfo
  {journal} {Contemp. Phys.},\ }\textbf {\bibinfo {volume} {48}},\ \bibinfo
  {pages} {13} (\bibinfo {year} {2007})}\BibitemShut {NoStop}%
\bibitem [{\citenamefont {L{\"a}uchli}\ and\ \citenamefont
  {Kollath}(2008)}]{Lauchli:2008}%
  \BibitemOpen
  \bibfield  {author} {\bibinfo {author} {\bibfnamefont {A.~M.}\ \bibnamefont
  {L{\"a}uchli}}\ and\ \bibinfo {author} {\bibfnamefont {C.}~\bibnamefont
  {Kollath}},\ }\Doi {10.1088/1742-5468/2008/05/P05018} {\bibfield  {journal}
  {\bibinfo  {journal} {J. Stat. Mech.},\ \bibinfo {pages} {P05018}} (\bibinfo
  {year} {2008})}\BibitemShut {NoStop}%
\bibitem [{\citenamefont {Nachtergaele}\ \emph {et~al.}(2009)\citenamefont
  {Nachtergaele}, \citenamefont {Raz}, \citenamefont {Schlein},\ and\
  \citenamefont {Sims}}]{Nachtergaele:2009a}%
  \BibitemOpen
  \bibfield  {author} {\bibinfo {author} {\bibfnamefont {B.}~\bibnamefont
  {Nachtergaele}}, \bibinfo {author} {\bibfnamefont {H.}~\bibnamefont {Raz}},
  \bibinfo {author} {\bibfnamefont {B.}~\bibnamefont {Schlein}}, \ and\
  \bibinfo {author} {\bibfnamefont {R.}~\bibnamefont {Sims}},\ }\Doi
  {10.1007/s00220-008-0630-2} {\bibfield  {journal} {\bibinfo  {journal}
  {Commun. Math. Phys.},\ }\textbf {\bibinfo {volume} {286}},\ \bibinfo {pages}
  {1073} (\bibinfo {year} {2009})}\BibitemShut {NoStop}%
\bibitem [{\citenamefont {Cramer}\ \emph {et~al.}(2008)\citenamefont {Cramer},
  \citenamefont {Serafini},\ and\ \citenamefont {Eisert}}]{Cramer:2008c}%
  \BibitemOpen
  \bibfield  {author} {\bibinfo {author} {\bibfnamefont {M.}~\bibnamefont
  {Cramer}}, \bibinfo {author} {\bibfnamefont {A.}~\bibnamefont {Serafini}}, \
  and\ \bibinfo {author} {\bibfnamefont {J.}~\bibnamefont {Eisert}},\ }in\
  \href@noop {} {\emph {\bibinfo {booktitle} {Quantum Information and Many Body
  Quantum Systems}}},\ \bibinfo {series} {CRM}, Vol.~\bibinfo {volume} {8},\
  \bibinfo {organization} {Publications of the Scuola Normale Superiore}\
  (\bibinfo  {publisher} {Edizioni della Normale},\ \bibinfo {address} {Pisa},\
  \bibinfo {year} {2008})\ pp.\ \bibinfo {pages} {51--72}\BibitemShut {NoStop}%
\bibitem [{\citenamefont {Eisert}\ and\ \citenamefont
  {Gross}(2009)}]{Eisert:2009}%
  \BibitemOpen
  \bibfield  {author} {\bibinfo {author} {\bibfnamefont {J.}~\bibnamefont
  {Eisert}}\ and\ \bibinfo {author} {\bibfnamefont {D.}~\bibnamefont {Gross}},\
  }\Doi {10.1103/PhysRevLett.102.240501} {\bibfield  {journal} {\bibinfo
  {journal} {Phys. Rev. Lett.},\ }\textbf {\bibinfo {volume} {102}},\ \bibinfo
  {pages} {240501} (\bibinfo {year} {2009})}\BibitemShut {NoStop}%
\bibitem [{\citenamefont {Hastings}(2004)}]{Hastings:2004a}%
  \BibitemOpen
  \bibfield  {author} {\bibinfo {author} {\bibfnamefont {M.~B.}\ \bibnamefont
  {Hastings}},\ }\Doi {10.1103/PhysRevB.69.104431} {\bibfield  {journal}
  {\bibinfo  {journal} {Phys. Rev. B},\ }\textbf {\bibinfo {volume} {69}},\
  \bibinfo {pages} {104431} (\bibinfo {year} {2004})}\BibitemShut {NoStop}%
\bibitem [{\citenamefont {Nachtergaele}\ and\ \citenamefont
  {Sims}(2011)}]{Nachtergaele:2011}%
  \BibitemOpen
  \bibfield  {author} {\bibinfo {author} {\bibfnamefont {B.}~\bibnamefont
  {Nachtergaele}}\ and\ \bibinfo {author} {\bibfnamefont {R.}~\bibnamefont
  {Sims}},\ }\href@noop {} { (\bibinfo {year} {2011})},\ \Eprint
  {http://arxiv.org/abs/arXiv:1102.0835} {arXiv:1102.0835} \BibitemShut
  {NoStop}%
\bibitem [{\citenamefont {Nachtergaele}\ and\ \citenamefont
  {Sims}(2006)}]{Nachtergaele:2006b}%
  \BibitemOpen
  \bibfield  {author} {\bibinfo {author} {\bibfnamefont {B.}~\bibnamefont
  {Nachtergaele}}\ and\ \bibinfo {author} {\bibfnamefont {R.}~\bibnamefont
  {Sims}},\ }\Doi {10.1007/s00220-006-1556-1} {\bibfield  {journal} {\bibinfo
  {journal} {Commun. Math. Phys.},\ }\textbf {\bibinfo {volume} {265}},\
  \bibinfo {pages} {119} (\bibinfo {year} {2006})}\BibitemShut {NoStop}%
\bibitem [{\citenamefont {Hastings}\ and\ \citenamefont
  {Koma}(2006)}]{Hastings:2006}%
  \BibitemOpen
  \bibfield  {author} {\bibinfo {author} {\bibfnamefont {M.~B.}\ \bibnamefont
  {Hastings}}\ and\ \bibinfo {author} {\bibfnamefont {T.}~\bibnamefont
  {Koma}},\ }\Doi {10.1007/s00220-006-0030-4} {\bibfield  {journal} {\bibinfo
  {journal} {Commun. Math. Phys.},\ }\textbf {\bibinfo {volume} {265}},\
  \bibinfo {pages} {781} (\bibinfo {year} {2006})}\BibitemShut {NoStop}%
\bibitem [{\citenamefont {Eisert}\ \emph {et~al.}(2010)\citenamefont {Eisert},
  \citenamefont {Cramer},\ and\ \citenamefont {Plenio}}]{Eisert:2010}%
  \BibitemOpen
  \bibfield  {author} {\bibinfo {author} {\bibfnamefont {J.}~\bibnamefont
  {Eisert}}, \bibinfo {author} {\bibfnamefont {M.}~\bibnamefont {Cramer}}, \
  and\ \bibinfo {author} {\bibfnamefont {M.~B.}\ \bibnamefont {Plenio}},\ }\Doi
  {10.1103/RevModPhys.82.277} {\bibfield  {journal} {\bibinfo  {journal} {Rev.
  Mod. Phys.},\ }\textbf {\bibinfo {volume} {82}},\ \bibinfo {pages} {277}
  (\bibinfo {year} {2010})}\BibitemShut {NoStop}%
\bibitem [{\citenamefont {Bloch}\ \emph {et~al.}(2008)\citenamefont {Bloch},
  \citenamefont {Dalibard},\ and\ \citenamefont {Zwerger}}]{Bloch:2008}%
  \BibitemOpen
  \bibfield  {author} {\bibinfo {author} {\bibfnamefont {I.}~\bibnamefont
  {Bloch}}, \bibinfo {author} {\bibfnamefont {J.}~\bibnamefont {Dalibard}}, \
  and\ \bibinfo {author} {\bibfnamefont {W.}~\bibnamefont {Zwerger}},\ }\Doi
  {10.1103/RevModPhys.80.885} {\bibfield  {journal} {\bibinfo  {journal} {Rev.
  Mod. Phys.},\ }\textbf {\bibinfo {volume} {80}},\ \bibinfo {pages} {885}
  (\bibinfo {year} {2008})}\BibitemShut {NoStop}%
\bibitem [{\citenamefont {Bakr}\ \emph {et~al.}(2009)\citenamefont {Bakr},
  \citenamefont {Gillen}, \citenamefont {Peng}, \citenamefont {Folling},\ and\
  \citenamefont {Greiner}}]{Bakr:2009}%
  \BibitemOpen
  \bibfield  {author} {\bibinfo {author} {\bibfnamefont {W.~S.}\ \bibnamefont
  {Bakr}}, \bibinfo {author} {\bibfnamefont {J.~I.}\ \bibnamefont {Gillen}},
  \bibinfo {author} {\bibfnamefont {A.}~\bibnamefont {Peng}}, \bibinfo {author}
  {\bibfnamefont {S.}~\bibnamefont {Folling}}, \ and\ \bibinfo {author}
  {\bibfnamefont {M.}~\bibnamefont {Greiner}},\ }\Doi {10.1038/nature08482}
  {\bibfield  {journal} {\bibinfo  {journal} {Nature},\ }\textbf {\bibinfo
  {volume} {462}},\ \bibinfo {pages} {74} (\bibinfo {year} {2009})}\BibitemShut
  {NoStop}%
\bibitem [{\citenamefont {Sherson}\ \emph {et~al.}(2010)\citenamefont
  {Sherson}, \citenamefont {Weitenberg}, \citenamefont {Endres}, \citenamefont
  {Cheneau}, \citenamefont {Bloch},\ and\ \citenamefont {Kuhr}}]{Sherson:2010}%
  \BibitemOpen
  \bibfield  {author} {\bibinfo {author} {\bibfnamefont {J.~F.}\ \bibnamefont
  {Sherson}}, \bibinfo {author} {\bibfnamefont {C.}~\bibnamefont {Weitenberg}},
  \bibinfo {author} {\bibfnamefont {M.}~\bibnamefont {Endres}}, \bibinfo
  {author} {\bibfnamefont {M.}~\bibnamefont {Cheneau}}, \bibinfo {author}
  {\bibfnamefont {I.}~\bibnamefont {Bloch}}, \ and\ \bibinfo {author}
  {\bibfnamefont {S.}~\bibnamefont {Kuhr}},\ }\Doi {10.1038/nature09378}
  {\bibfield  {journal} {\bibinfo  {journal} {Nature},\ }\textbf {\bibinfo
  {volume} {467}},\ \bibinfo {pages} {68} (\bibinfo {year} {2010})}\BibitemShut
  {NoStop}%
\bibitem [{\citenamefont {Fisher}\ \emph {et~al.}(1989)\citenamefont {Fisher},
  \citenamefont {Weichman}, \citenamefont {Grinstein},\ and\ \citenamefont
  {Fisher}}]{Fisher:1989}%
  \BibitemOpen
  \bibfield  {author} {\bibinfo {author} {\bibfnamefont {M.~P.~A.}\
  \bibnamefont {Fisher}}, \bibinfo {author} {\bibfnamefont {P.~B.}\
  \bibnamefont {Weichman}}, \bibinfo {author} {\bibfnamefont {G.}~\bibnamefont
  {Grinstein}}, \ and\ \bibinfo {author} {\bibfnamefont {D.~S.}\ \bibnamefont
  {Fisher}},\ }\Doi {10.1103/PhysRevB.40.546} {\bibfield  {journal} {\bibinfo
  {journal} {Phys. Rev. B},\ }\textbf {\bibinfo {volume} {40}},\ \bibinfo
  {pages} {546} (\bibinfo {year} {1989})}\BibitemShut {NoStop}%
\bibitem [{\citenamefont {Jaksch}\ \emph {et~al.}(1998)\citenamefont {Jaksch},
  \citenamefont {Bruder}, \citenamefont {Cirac}, \citenamefont {Gardiner},\
  and\ \citenamefont {Zoller}}]{Jaksch:1998a}%
  \BibitemOpen
  \bibfield  {author} {\bibinfo {author} {\bibfnamefont {D.}~\bibnamefont
  {Jaksch}}, \bibinfo {author} {\bibfnamefont {C.}~\bibnamefont {Bruder}},
  \bibinfo {author} {\bibfnamefont {J.~I.}\ \bibnamefont {Cirac}}, \bibinfo
  {author} {\bibfnamefont {C.~W.}\ \bibnamefont {Gardiner}}, \ and\ \bibinfo
  {author} {\bibfnamefont {P.}~\bibnamefont {Zoller}},\ }\Doi
  {10.1103/PhysRevLett.81.3108} {\bibfield  {journal} {\bibinfo  {journal}
  {Phys. Rev. Lett.},\ }\textbf {\bibinfo {volume} {81}},\ \bibinfo {pages}
  {3108} (\bibinfo {year} {1998})}\BibitemShut {NoStop}%
\bibitem [{\citenamefont {K{\"u}hner}\ \emph {et~al.}(2000)\citenamefont
  {K{\"u}hner}, \citenamefont {White},\ and\ \citenamefont
  {Monien}}]{Kuhner:2000}%
  \BibitemOpen
  \bibfield  {author} {\bibinfo {author} {\bibfnamefont {T.~D.}\ \bibnamefont
  {K{\"u}hner}}, \bibinfo {author} {\bibfnamefont {S.~R.}\ \bibnamefont
  {White}}, \ and\ \bibinfo {author} {\bibfnamefont {H.}~\bibnamefont
  {Monien}},\ }\Doi {10.1103/PhysRevB.61.12474} {\bibfield  {journal} {\bibinfo
   {journal} {Phys. Rev. B},\ }\textbf {\bibinfo {volume} {61}},\ \bibinfo
  {pages} {12474} (\bibinfo {year} {2000})}\BibitemShut {NoStop}%
\bibitem [{\citenamefont {Endres}\ \emph {et~al.}(2011)\citenamefont {Endres},
  \citenamefont {Cheneau}, \citenamefont {Fukuhara}, \citenamefont
  {Weitenberg}, \citenamefont {Schau{\ss}}, \citenamefont {Gross},
  \citenamefont {Mazza}, \citenamefont {Ba{\~n}uls}, \citenamefont {Pollet},
  \citenamefont {Bloch},\ and\ \citenamefont {Kuhr}}]{Endres:2011}%
  \BibitemOpen
  \bibfield  {author} {\bibinfo {author} {\bibfnamefont {M.}~\bibnamefont
  {Endres}}, \bibinfo {author} {\bibfnamefont {M.}~\bibnamefont {Cheneau}},
  \bibinfo {author} {\bibfnamefont {T.}~\bibnamefont {Fukuhara}}, \bibinfo
  {author} {\bibfnamefont {C.}~\bibnamefont {Weitenberg}}, \bibinfo {author}
  {\bibfnamefont {P.}~\bibnamefont {Schau{\ss}}}, \bibinfo {author}
  {\bibfnamefont {C.}~\bibnamefont {Gross}}, \bibinfo {author} {\bibfnamefont
  {L.}~\bibnamefont {Mazza}}, \bibinfo {author} {\bibfnamefont {M.~C.}\
  \bibnamefont {Ba{\~n}uls}}, \bibinfo {author} {\bibfnamefont
  {L.}~\bibnamefont {Pollet}}, \bibinfo {author} {\bibfnamefont
  {I.}~\bibnamefont {Bloch}}, \ and\ \bibinfo {author} {\bibfnamefont
  {S.}~\bibnamefont {Kuhr}},\ }\Doi {10.1126/science.1209284} {\bibfield
  {journal} {\bibinfo  {journal} {Science},\ }\textbf {\bibinfo {volume}
  {334}},\ \bibinfo {pages} {200} (\bibinfo {year} {2011})}\BibitemShut
  {NoStop}%
\bibitem [{\citenamefont {Daley}\ \emph {et~al.}(2004)\citenamefont {Daley},
  \citenamefont {Kollath}, \citenamefont {Schollw{\"o}ck},\ and\ \citenamefont
  {Vidal}}]{Daley:2004}%
  \BibitemOpen
  \bibfield  {author} {\bibinfo {author} {\bibfnamefont {A.~J.}\ \bibnamefont
  {Daley}}, \bibinfo {author} {\bibfnamefont {C.}~\bibnamefont {Kollath}},
  \bibinfo {author} {\bibfnamefont {U.}~\bibnamefont {Schollw{\"o}ck}}, \ and\
  \bibinfo {author} {\bibfnamefont {G.}~\bibnamefont {Vidal}},\ }\Doi
  {10.1088/1742-5468/2004/04/P04005} {\bibfield  {journal} {\bibinfo  {journal}
  {J. Stat. Mech.},\ }\textbf {\bibinfo {volume} {2004}},\ \bibinfo {pages}
  {P04005} (\bibinfo {year} {2004})}\BibitemShut {NoStop}%
\bibitem [{\citenamefont {White}\ and\ \citenamefont
  {Feiguin}(2004)}]{White:2004}%
  \BibitemOpen
  \bibfield  {author} {\bibinfo {author} {\bibfnamefont {S.~R.}\ \bibnamefont
  {White}}\ and\ \bibinfo {author} {\bibfnamefont {A.~E.}\ \bibnamefont
  {Feiguin}},\ }\Doi {10.1103/PhysRevLett.93.076401} {\bibfield  {journal}
  {\bibinfo  {journal} {Phys. Rev. Lett.},\ }\textbf {\bibinfo {volume} {93}},\
  \bibinfo {pages} {076401} (\bibinfo {year} {2004})}\BibitemShut {NoStop}%
\bibitem [{\citenamefont {Kollath}\ \emph {et~al.}(2005)\citenamefont
  {Kollath}, \citenamefont {Schollw{\"o}ck}, \citenamefont {von Delft},\ and\
  \citenamefont {Zwerger}}]{Kollath:2005b}%
  \BibitemOpen
  \bibfield  {author} {\bibinfo {author} {\bibfnamefont {C.}~\bibnamefont
  {Kollath}}, \bibinfo {author} {\bibfnamefont {U.}~\bibnamefont
  {Schollw{\"o}ck}}, \bibinfo {author} {\bibfnamefont {J.}~\bibnamefont {von
  Delft}}, \ and\ \bibinfo {author} {\bibfnamefont {W.}~\bibnamefont
  {Zwerger}},\ }\Doi {10.1103/PhysRevA.71.053606} {\bibfield  {journal}
  {\bibinfo  {journal} {Phys. Rev. A},\ }\textbf {\bibinfo {volume} {71}},\
  \bibinfo {pages} {053606} (\bibinfo {year} {2005})}\BibitemShut {NoStop}%
\bibitem [{\citenamefont {Kibble}(1976)}]{Kibble:1976}%
  \BibitemOpen
  \bibfield  {author} {\bibinfo {author} {\bibfnamefont {T.~W.~B.}\
  \bibnamefont {Kibble}},\ }\Doi {10.1088/0305-4470/9/8/029} {\bibfield
  {journal} {\bibinfo  {journal} {J. Phys. A-Math. Gen.},\ }\textbf {\bibinfo
  {volume} {9}},\ \bibinfo {pages} {1387} (\bibinfo {year} {1976})}\BibitemShut
  {NoStop}%
\bibitem [{\citenamefont {Zurek}(1985)}]{Zurek:1985}%
  \BibitemOpen
  \bibfield  {author} {\bibinfo {author} {\bibfnamefont {W.~H.}\ \bibnamefont
  {Zurek}},\ }\Doi {10.1038/317505a0} {\bibfield  {journal} {\bibinfo
  {journal} {Nature},\ }\textbf {\bibinfo {volume} {317}},\ \bibinfo {pages}
  {505} (\bibinfo {year} {1985})}\BibitemShut {NoStop}%
\bibitem [{\citenamefont {Batista}\ and\ \citenamefont
  {Ortiz}(2001)}]{Batista:2001}%
  \BibitemOpen
  \bibfield  {author} {\bibinfo {author} {\bibfnamefont {C.~D.}\ \bibnamefont
  {Batista}}\ and\ \bibinfo {author} {\bibfnamefont {G.}~\bibnamefont
  {Ortiz}},\ }\Doi {10.1103/PhysRevLett.86.1082} {\bibfield  {journal}
  {\bibinfo  {journal} {Phys. Rev. Lett.},\ }\textbf {\bibinfo {volume} {86}},\
  \bibinfo {pages} {1082} (\bibinfo {year} {2001})}\BibitemShut {NoStop}%
\bibitem [{\citenamefont {Huber}\ \emph {et~al.}(2007)\citenamefont {Huber},
  \citenamefont {Altman}, \citenamefont {B{\"u}chler},\ and\ \citenamefont
  {Blatter}}]{Huber:2007}%
  \BibitemOpen
  \bibfield  {author} {\bibinfo {author} {\bibfnamefont {S.~D.}\ \bibnamefont
  {Huber}}, \bibinfo {author} {\bibfnamefont {E.}~\bibnamefont {Altman}},
  \bibinfo {author} {\bibfnamefont {H.~P.}\ \bibnamefont {B{\"u}chler}}, \ and\
  \bibinfo {author} {\bibfnamefont {G.}~\bibnamefont {Blatter}},\ }\Doi
  {10.1103/PhysRevB.75.085106} {\bibfield  {journal} {\bibinfo  {journal}
  {Phys. Rev. B},\ }\textbf {\bibinfo {volume} {75}},\ \bibinfo {pages}
  {085106} (\bibinfo {year} {2007})}\BibitemShut {NoStop}%
\bibitem [{\citenamefont {Altman}\ and\ \citenamefont
  {Auerbach}(2002)}]{Altman:2002}%
  \BibitemOpen
  \bibfield  {author} {\bibinfo {author} {\bibfnamefont {E.}~\bibnamefont
  {Altman}}\ and\ \bibinfo {author} {\bibfnamefont {A.}~\bibnamefont
  {Auerbach}},\ }\Doi {10.1103/PhysRevLett.89.250404} {\bibfield  {journal}
  {\bibinfo  {journal} {Phys. Rev. Lett.},\ }\textbf {\bibinfo {volume} {89}},\
  \bibinfo {pages} {250404} (\bibinfo {year} {2002})}\BibitemShut {NoStop}%
\bibitem [{\citenamefont {Schollw{\"o}ck}(2011)}]{Schollwock:2011}%
  \BibitemOpen
  \bibfield  {author} {\bibinfo {author} {\bibfnamefont {U.}~\bibnamefont
  {Schollw{\"o}ck}},\ }\Doi {10.1016/j.aop.2010.09.012} {\bibfield  {journal}
  {\bibinfo  {journal} {Ann. Phys. (N.Y.)},\ }\textbf {\bibinfo {volume}
  {326}},\ \bibinfo {pages} {96} (\bibinfo {year} {2011})}\BibitemShut
  {NoStop}%
\bibitem [{\citenamefont {Vidal}(2007)}]{Vidal:2007}%
  \BibitemOpen
  \bibfield  {author} {\bibinfo {author} {\bibfnamefont {G.}~\bibnamefont
  {Vidal}},\ }\Doi {10.1103/PhysRevLett.98.070201} {\bibfield  {journal}
  {\bibinfo  {journal} {Phys. Rev. Lett.},\ }\textbf {\bibinfo {volume} {98}},\
  \bibinfo {pages} {070201} (\bibinfo {year} {2007})}\BibitemShut {NoStop}%
\bibitem [{\citenamefont {McCulloch}(2008)}]{McCulloch:2008}%
  \BibitemOpen
  \bibfield  {author} {\bibinfo {author} {\bibfnamefont {I.~P.}\ \bibnamefont
  {McCulloch}},\ }\href@noop {} { (\bibinfo {year} {2008})},\ \Eprint
  {http://arxiv.org/abs/arXiv:0804.2509} {arXiv:0804.2509} \BibitemShut
  {NoStop}%
\bibitem [{\citenamefont {White}(1992)}]{White:1992}%
  \BibitemOpen
  \bibfield  {author} {\bibinfo {author} {\bibfnamefont {S.~R.}\ \bibnamefont
  {White}},\ }\Doi {10.1103/PhysRevLett.69.2863} {\bibfield  {journal}
  {\bibinfo  {journal} {Phys. Rev. Lett.},\ }\textbf {\bibinfo {volume} {69}},\
  \bibinfo {pages} {2863} (\bibinfo {year} {1992})}\BibitemShut {NoStop}%
\end{thebibliography}%

\newpage%
\section*{Appendix}

\subsection*{Quenches to \mbox{U/J\,=\,5.0} and 7.0}

\begin{figure}[b]
  \includegraphics[width=1.0\columnwidth]{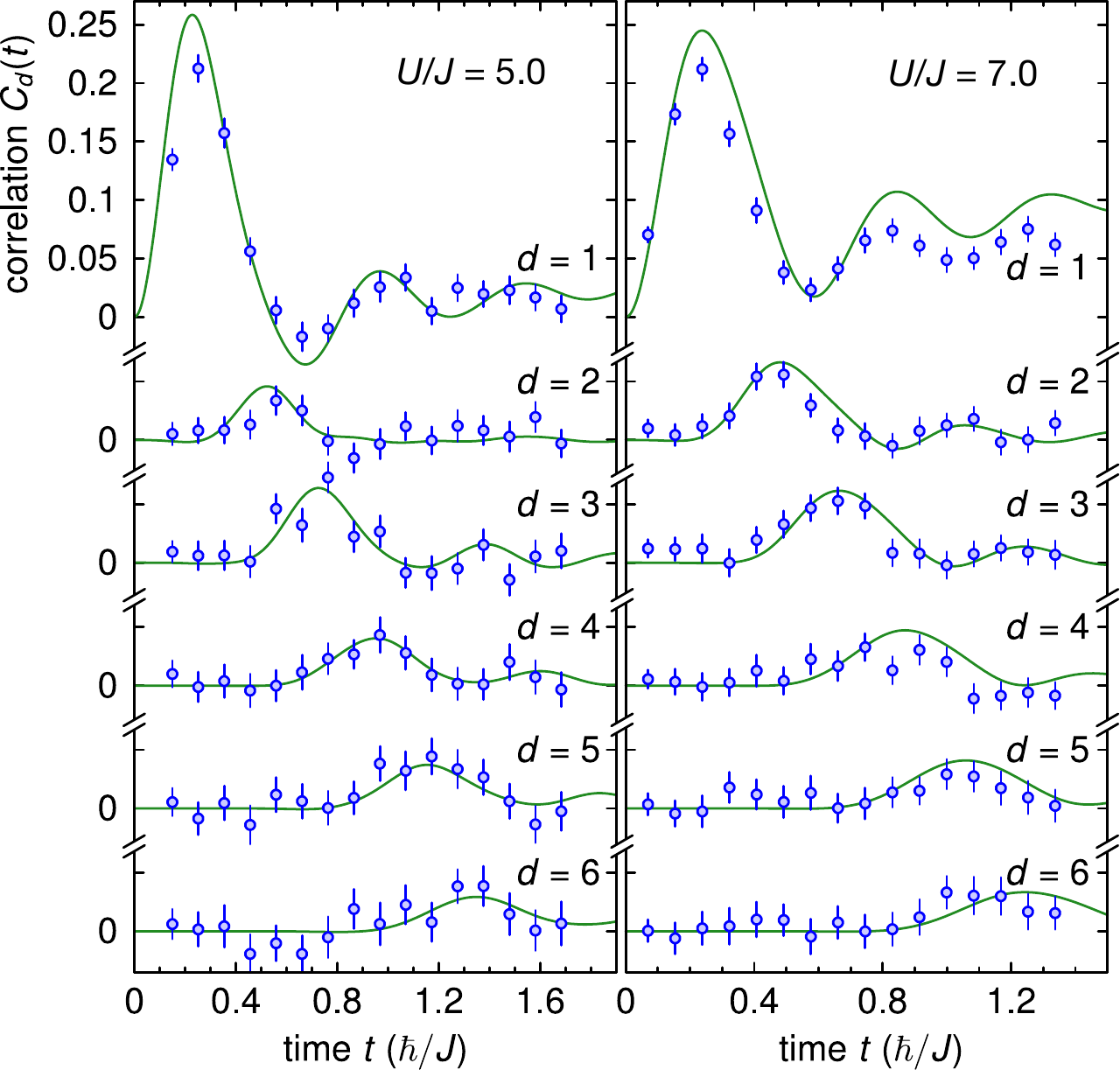}
  \caption{\textbf{Time evolution of the two-point parity correlations.} Left panel: quench to $U/J
    = 5.0(2)$. Right panel: quench to $U/J = 7.0(3)$. The circles indicate the correlations measured
    experimentally and the line is derived from the numerical simulations for an infinite,
    homogeneous system at zero temperature. The experimental and numerical values were obtained in
    the same way as described in the legend of Fig.~\ref{fig:2} and in the Methods Summary section.}
  \label{fig:S1}
\end{figure}

We also recorded the time evolution of the two-point parity correlations (2) after quenches to
$U/J=5.0(2)$ and 7.0(3), and compared the experimental results to DMRG simulations of an infinite,
homogeneous system at zero temperature (Fig.~\ref{fig:S1}). The experimental sequence was identical
to the one we used for the quench to $U/J = 9.0(3)$, apart from the different end point of the
quench. The data presented here are those used in Fig.~\ref{fig:3}.

\subsection*{Quasiparticle model}
\label{sec:ferm-quas-model}

\def \double{{\ket{\begin{smallmatrix} \bullet \\ \bullet \end{smallmatrix}}}}%
\def \hole{{\ket{\begin{smallmatrix} \circ \\ \circ \end{smallmatrix}}}}%
\def \single{{\ket{\begin{smallmatrix} \circ \\ \bullet \end{smallmatrix}}}}%
\def \vac{\ket{\begin{smallmatrix} \circ \\ \bullet \end{smallmatrix} \begin{smallmatrix} \circ \\
      \bullet \end{smallmatrix} \cdots}}

In the Bose--Hubbard model, bosonic atoms in an optical lattice are confined to a single Bloch band
and obey the Hamiltonian
\begin{equation}
  \hat H = \sum_{j} \Big\{ -J \,\big( \hat a^{\dagger}_{j} \, \hat a_{j+1} + \text{h.\,c.} \big) +
  \frac{U}{2} \hat n_{j} (\hat n_{j}-1) \Big\} \; ,
  \label{eq:3}
\end{equation}
where $\hat a_{j}$ and $\hat a^{\dagger}_{j}$ represent the annihilation and creation operator of an
atom at site $j$ and $\hat n_{j} = \hat a^{\dagger}_{j} \hat a_{j}$ counts the number of atoms at
that site. The model is entirely parametrised by the effective interaction strength $U/J$.

In order to analytically treat the time evolution of correlations after a sudden decrease of $U/J$,
we developed a novel approach based on fermionized quasiparticles. The initial state being close to
a Fock state with one atom per lattice site, an effective description of the evolution at
sufficiently large final interaction strengths can be obtained within a local basis formed by empty
states, $\hole_j$, singly occupied states, $\single_j$, and doubly occupied states,
$\double_j$. Using generalised Jordan--Wigner transformations \cite{Batista:2001}, we then
introduced fermionic creation operators for the excess particles, \mbox{$\hat
  d_j^\dagger\single_j\to\double_j$}, and the holes, \mbox{$\hat h_j^\dagger\single_j\to\hole_j$},
as well as the corresponding annihilation operators. Within the truncated Hilbert space, the
original Hamiltonian \eqref{eq:3} can be exactly written in terms of these operators:
\begin{multline}
  \label{eq:4}
  \hat H = \parbox[t][\ht\strutbox]{\widthof{$\displaystyle \sum_j $}}{$\displaystyle \sum_j$} \;
  \hat{\mathcal{P}} \; \Big\{ -2J \, \hat d_{j}^\dagger \, \hat d_{j+1} - J \, \hat h^\dagger_{j+1}
  \, \hat h_{j} \\%
  - J\sqrt{2} \big( \hat d_{j}^\dagger \, \hat h^\dagger_{j+1} - \hat h_{j} \, \hat d_{j+1} \big)
  +\text{h.\,c} \\%
  + \frac{U}{2} \big( \hat n_{\text{d},j} \, + \hat n_{\text{h},j}\big) \Big\} \; \hat{\mathcal{P}}
  \; ,
\end{multline}
with $\hat n_{\text{d},j}=\hat d^{\dagger}_{j} \hat d_{j}$ and $ \hat n_{\text{h},j}=\hat
h^{\dagger}_{j} \hat h_{j}$. The complexity of the model is hidden in the projector
$\hat{\mathcal{P}} = \prod_j (\hat I-\hat n_{\text{d},j}\hat n_{\text{h},j})$ that eliminates the
unphysical situation of having an excess particle and a hole at the same site ($\hat I$ is the
identity operator). As multiple occupancies of equal species are naturally avoided due to their
statistics, one still obtains a good description of the system when setting $\hat{\mathcal{P}}
\rightarrow \hat I$, provided the density of excitations $\avg{ \hat n_{\text{d},j}(t)+\hat
  n_{\text{h},j}(t)}$ remains low. This is in contrast to the usual bosonic representations
\cite{Huber:2007, Altman:2002}.

The Hamiltonian \eqref{eq:4} with $\hat{\mathcal{P}} \rightarrow \hat I$ is quadratic and can be
diagonalised by a Bogolyubov transformation. The eigenmodes are \textit{doublons} and
\textit{holons} with well defined momentum $k$:
\begin{align}
  \label{eq:5}
  \hat \gamma_{\text{d},k}^\dagger & = u(k) \, \hat d^{\dagger}_k + v(k) \, \hat h_{-k} \; , \\
  \label{eq:6}
  \hat \gamma_{\text{h},-k}^\dagger & = u(k) \, \hat h^\dagger_{-k} - v(k) \, \hat d_{k} \; ,
\end{align}
with
\begin{multline}
  \label{eq:7}
  u(k) = \cos[\theta(k)/2] \; , \quad  v(k) = i\sin[\theta(k)/2] \\
  \text{and} \quad \theta(k) =
  \text{atan}\left[\frac{\sqrt{32}J\sin(ka_{\text{lat}})}{U-6J\cos(ka_{\text{lat}})} \right] \; .
\end{multline}
Their respective eigenenergies are given by
\begin{gather}
  \label{eq:8}
  \begin{multlined}[b][0.85\columnwidth]
    \epsilon_{\text{d}}(k) = - J\cos(ka_{\text{lat}}) \\
    \shoveleft{+\frac{1}{2}\sqrt{[U-6J\cos(ka_{\text{lat}})]^2 + 32 J^2 \sin^2(ka_{\text{lat}})}}
    \,\, ,
  \end{multlined} \\
  \label{eq:9}
  \begin{multlined}[b][0.85\columnwidth]
    \epsilon_{\text{h}}(k) = J\cos(ka_{\text{lat}}) \\
    \shoveleft{+\frac{1}{2}\sqrt{[U-6J\cos(ka_{\text{lat}})]^2 + 32 J^2 \sin^2(ka_{\text{lat}})}}
    \,\, .
  \end{multlined}
\end{gather}

Within this model, the initial state $\ket{\psi_0}$ evolves in time according to:
\begin{align}
  \label{eq:10}
  \ket{\psi(t)} & = e^{-i \hat Ht/\hbar} \, \ket{\psi_0} \\
  \label{eq:11}
  & \hspace{-1.5pt}
  \begin{multlined}[b]
    = \parbox[t][\ht\strutbox]{0.65\columnwidth}{$\displaystyle \prod_{k} \Big\{ \bar{u}(k) -
      \bar{v}(k) \, e^{-i[\epsilon_{\text{d}}(k) + \epsilon_{\text{h}}(-k) ] t/\hbar} $} \\
    \cdot \; \hat \gamma^\dagger_{\text{d},k} \, \hat \gamma_{\text{h},-k}^{\dagger} \Big\}
    \ket{\text{vac}} \; ,
  \end{multlined}
\end{align}
with
\begin{align}
  \bar{u}(k) & = u(k) u_{0}(k) - v(k) v_{0}(k) \; , \\
  \bar{v}(k) & = v(k) u_{0}(k) - u(k) v_{0}(k) \; .
\end{align}
Here the subscript ``0'' denotes quantities corresponding to the initial interaction strength,
whereas no label is used for the quantities corresponding to the final interaction
strength. Further, $\ket{\text{vac}}$ represents the quasiparticle vacuum at the final interaction
strength ($\hat \gamma_{\text{d},k}\ket{\text{vac}} = \hat
\gamma_{\text{h},k}\ket{\text{vac}}=0$). Equation \eqref{eq:10} describes wave packets of entangled
quasiparticle pairs that travel in opposite directions with different velocities. One can extract a
maximal velocity for the spreading of the correlations from the dispersion relation of a
quasiparticle pair (black line in Fig.~\ref{fig:3}b):
\begin{equation}
  \label{eq:12}
  v_{\text{max}} = \frac{1}{\hbar} \max_{k} \left\{ \frac{\text{d}}{\text{d}k} \big|
    \epsilon_{\text{d}}(k) + \epsilon_{\text{h}}(k) \big| \right\} \; .
\end{equation}

\begin{figure}
  \includegraphics[width=1.0\columnwidth]{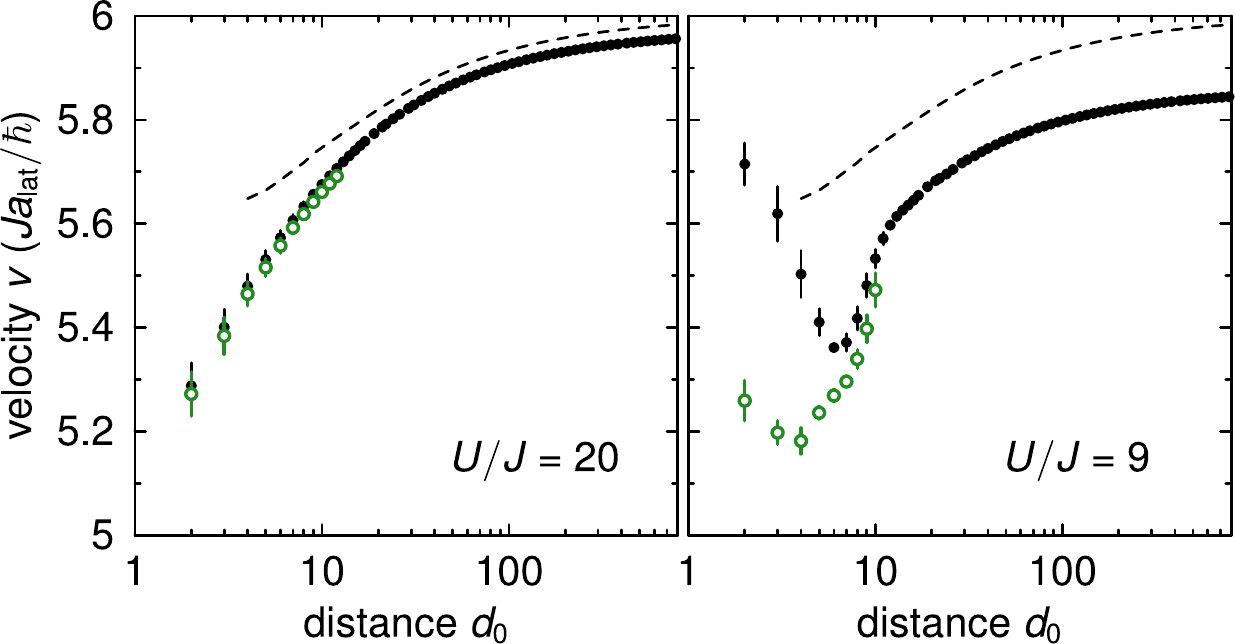}
  \caption{\textbf{Instantaneous propagation velocity}. We compare the instantaneous velocity
    $v(d_{0})$ predicted by our analytical model (points) with the one derived from numerical
    simulations (circles). The agreement is excellent at $U/J=20$ (left panel) and qualitatively
    good at $U/J=9$ (right panel). Error bars denote the 68\,\% confidence interval of the fit. The
    dashed line represents the asymptotic expression \eqref{eq:13}. Arrows point to the maximum
    velocity $v_{\text{max}}$ at the given interaction strength. The signal position was obtained
    using the procedure described in the Methods Summary section for the numerical simulations.}
  \label{fig:S2}
\end{figure}

Additionally, we derived from equation \eqref{eq:10} the time evolution of the two-point parity
correlations. In agreement with DMRG simulations, it displays a clear positive signal, the position
of which increases with time (Fig.~\ref{fig:2}). We extracted the instantaneous propagation velocity
$v(d_0)$ from a linear fit through the signal positions $d_{0} \leq d \leq d_{0}+4$
(Fig.~\ref{fig:S2}). The case $d_{0}=2$ corresponds to the data shown in Fig.~\ref{fig:3}. At short
distances, we find velocities about $10\,\%$ smaller than $v_{\text{max}}$, in good agreement with
the velocities measured experimentally. At large distances, the velocity converges algebraically to
$v_{\text{max}}$. This latter behaviour can be understood from the expansion of the correlation
functions to first order in $J/U$, which can be expressed in terms of the Airy function $\Ai(x)$:
\begin{equation}
  \label{eq:13}
  C_{d}(t) \stackrel{d \gg 1\vphantom{\big(}}{\simeq} \left[\frac{2^{7/3} d^{2/3} \hbar}{3\,U t}
    \, \Ai\left[ (2/d)^{1/3}(d-6Jt/\hbar) \right] \right]^2 \; .
\end{equation}

We checked the validity of our model of freely propagating fermionic quasiparticles by comparing it
with DMRG simulations. The propagation velocity of the two-point parity correlations is very
accurate in the strongly interacting limit (e.g. $U/J=20$), and remains in fairly good agreement
down to $U/J =9$, where the quasiparticle density is about $0.1$ per site (Fig.~\ref{fig:S2}). At
lower interaction strengths, we found that the Gutzwiller approximation breaks down. Nevertheless,
the experiment and the simulations show that the spreading of correlations is still characterised by
a well defined velocity, for which $v_{\text{max}}$ remains a relevant upper bound. We verified
numerically that the truncation of the local Hilbert space to three states is reasonable down to
$U/J \simeq 6$.

\subsection*{Calibration of the lattice depth}
We calibrated the lattice depths by performing amplitude modulation spectroscopy of the transition
between the zeroth and second Bloch band on a \oned degenerate gas for the $x$- and $y$-axes, and on
a \twod degenerate gas for the $z$-axis. We estimate the calibration uncertainty to be 1--2\%.

\subsection*{Bose--Hubbard parameters}

For a given lattice depth $V$, we calculated the tunnel coupling and the on-site interaction of the
Bose--Hubard model \eqref{eq:3} in the single-particle picture using their expressions as overlap
integrals of the Wannier functions. In Table~\ref{tab:1}, we provide the values of $V$, $J$ and $U$
for the effective interaction strengths mentioned in the main text.

\begin{table}
  \renewcommand{\arraystretch}{1.2}
  \begin{tabular}{|>{\centering} m{1cm}||>{\centering} m{1.5cm}|>{\centering} m{1.5cm}|>{\centering}m{1.5cm}|}
    \hline
    $U/J$ & $V$ (\Er) &  $J/\hbar~(\text{s}^{-1})$ & $U/h~(\text{Hz})$ \tabularnewline
    \hline
    40(2)  & 13.5(2) & 113(5) & 724(4) \tabularnewline
    9.0(3) & 7.70(11) & 422(11) & 607(3) \tabularnewline
    7.0(3) & 6.85(11) & 522(14) & 584(3) \tabularnewline
    5.0(2) & 5.75(13) & 691(23) & 550(4) \tabularnewline
    \hline
  \end{tabular}
  \caption{Lattice depth and Bose--Hubbard parameters for the data reported in the main text. The
    lattice depth in the perpendicular axes is $20.0(5)$\,\Er. The uncertainties on $J$, $U$ and
    $U/J$ follow from the calibration uncertainty on the lattice depth (see Methods Summary).}
  \label{tab:1}
\end{table}

\subsection*{Ramp of the lattice depth for the quench}

The ramp of the lattice depth for the quench follows the functional form:
\begin{equation}
  \label{eq:14}
  V(t) = V_{0} + (V-V_{0}) \, \frac{e^{-t/\tau}-1}{e^{-T/\tau}-1} \; .
\end{equation}
Here $V_{0} = 13.5\,\Er$ is the initial lattice depth, $T$ is the duration of the ramp and $\tau$
determines the rate at which the lattice depth is decreased. These parameters have to be chosen such
that the ramp is fast compared to the time scale of the dynamics following the quench, which is
given by $\hbar/J$, but adiabatic with regard to transitions to higher Bloch bands. The latter
condition requires the parameter $\Gamma = (\hbar/\Delta^{2})|\text{d}\Delta/\text{d}t|$ to be much
smaller than 1, where $\Delta$ is the energy gap between the two Bloch bands considered. For each
quench, we have chosen the ramp parameters such that $T$ is the shortest ramp duration compatible
with $\Gamma < 0.3$ (Table~\ref{tab:2}).

Numerical simulations showed that the origin $t=0$ of the time evolution can be defined as the
moment when the effective interaction strength reaches the value $U/J \simeq 17$. We used the same
phenomenological criterion to locate the moment at which the dynamics stops when raising the lattice
depth to $\sim 80\,\Er$.

\begin{table}
  \renewcommand{\arraystretch}{1.2}
  \begin{tabular}{|>{\centering} m{1cm}||>{\centering} m{1.2cm}|>{\centering} m{1.2cm}|}
    \hline
    $U/J$ & $T~(\text{\textmu s})$ & $\tau~(\text{\textmu s})$ \tabularnewline
    \hline
    9.0 & 100 & 130 \tabularnewline
    7.0 & 100 & 120 \tabularnewline
    5.0 & 150  & 160 \tabularnewline
    \hline
  \end{tabular}
  \caption{Lattice depth ramp parameters used for the quenches reported in the main text, as
    defined in equation \eqref{eq:14}.}
  \label{tab:2}
\end{table}

\subsection*{Determination of the time of the correlation peak}

We extracted the time of the maximum of the correlation signal as a function of the distance, for
both the experiment and the theory, by fitting an offset-free gaussian profile to the traces
$C_{d}(t)$. For the numerical data, we filtered out frequency components above $3\,J/\hbar$ prior to
the fit in order to isolate the envelope of the signal. For the experimental data, we fixed the
width of the gaussian profile to the value obtained from the numerical data, keeping only the
amplitude and time as free parameters.

\bigskip
\subsection*{Numerical simulations}

The numerical simulations relied on matrix product states \cite{Schollwock:2011} (MPS) to represent
infinite homogenous systems \cite{Vidal:2007, McCulloch:2008}. Initial states were obtained using
the DMRG algorithm \cite{White:1992}. For the time evolution, we used a second-order Suzuki--Trotter
decom\-po\-si\-tion \cite{White:2004, Daley:2004}. We achieved quasi-exact results on the relevant
time scale $tJ/\hbar \sim 2$ by choosing a small enough Trotter time step ($\sim 0.002$) and
retaining a few thousand states.

\end{document}